**Pressure Evolution of Atomic Volume Systematics in Transition Metals**

Masaaki GESHI[1] and Yuichi AKAHAMA[2,a)]

[1] $R^3$ Institute for Newly-Emerging Science Design, Osaka University,

1-3, Machikaneyama, Toyonaka, Osaka, Japan

[2] Department of Material Science, Graduate School of Science, University of Hyogo, 3-2-1 Kamigohri,

Hyogo 678-1297, Japan

[a]Author to whom correspondence should be addressed akahama@sci.u-hyogo.ac.jp

## ABSTRACT

We investigated the evolution of the well-known parabolic dependence of atomic volume on atomic number in transition metals under extreme compression at pressures up to 400 GPa using density functional theory calculations. Our results reveal that the ambient-pressure parabolic trend transforms into a characteristic cubic-like behavior at high pressures. This evolution is attributed to the higher compressibility of bcc transition metals associated with comparatively large increases in the total energy. The present findings are discussed in relation to previous experimental observations and first-principles calculations.

## 1. Introduction

Transition metals systematically exhibit a parabolic relationship between the atomic volume and atomic number at ambient pressure, thus reflecting their *d*-electron band structure.[1] Recently, we have reported based on experimental results in the multi-megabar pressure range, that this parabolic relationship changes significantly because group IV and V elements exhibit high compressibility owing to the stabilization of the bcc structure under high pressure.[2] The crystal structure of most transition metals, except for rare-earth elements, changes systematically with increasing *d*-band electron number, following the sequence bcc→hcp→fcc→hcp (see Supplementary Materials: Table S1).[3] Our recent theoretical study of the electronic band structures further suggested that the bcc phase of these transition metals is stabilized at pressures up to 1000 GPa.[4]

In this work, we investigate how the parabolic relationship between atomic volume and atomic number evolves under extreme compression. Using density functional theory (DFT) calculations for transition metals up to 400 GPa, we show that the ambient-pressure parabolic trend transforms into a characteristic cubic-like behavior.

## 2. Calculation Methods

DFT calculations were performed using Quantum ESPRESSO[5,6] for transition metals with bcc (Ti, V, Cr, Zr, Nb, Mo, Hf, Ta, and W) and fcc (Co, Ni, Cu, Rh, Pd, Ag, Ir, Pt, and Au) structures. The pressure-volume (*P*–*V*) relationship was obtained using the Perdew–Burke–Ernzerhof (PBE) exchange-correlation functional[7] and ultrasoft pseudopotentials.[8] The corresponding Brillouin zones were sampled with a 24 × 24 × 24 *k*-point mesh for both bcc and fcc structures, employing the Marzari–Vanderbilt smearing method with a broadening energy width of 0.02 Ry. For transition metals with a hcp structures (Ti, Mn, Fe, Co, Zn, Zr, Tc, Ru, Cd, Hf, Re, Os, and Hg), the *P*-*V* relationship was calculated using variable cell relaxation (vc-relax) to optimize the *c*/*a* ratio with a 24 × 24 × 18 *k*-point mesh. The cutoff energies were set to 1.5 to 3 times the recommended value, typically 80 Ry for the wave functions and 960 Ry for the charge density, and the convergence threshold was fixed at $1\times10^{-9}$. The *P*–*V* relationship was calculated up to 400 GPa by computing the stress and pressure at intervals of approximately 20 GPa, and the results were used to estimate the equation of state (EOS).

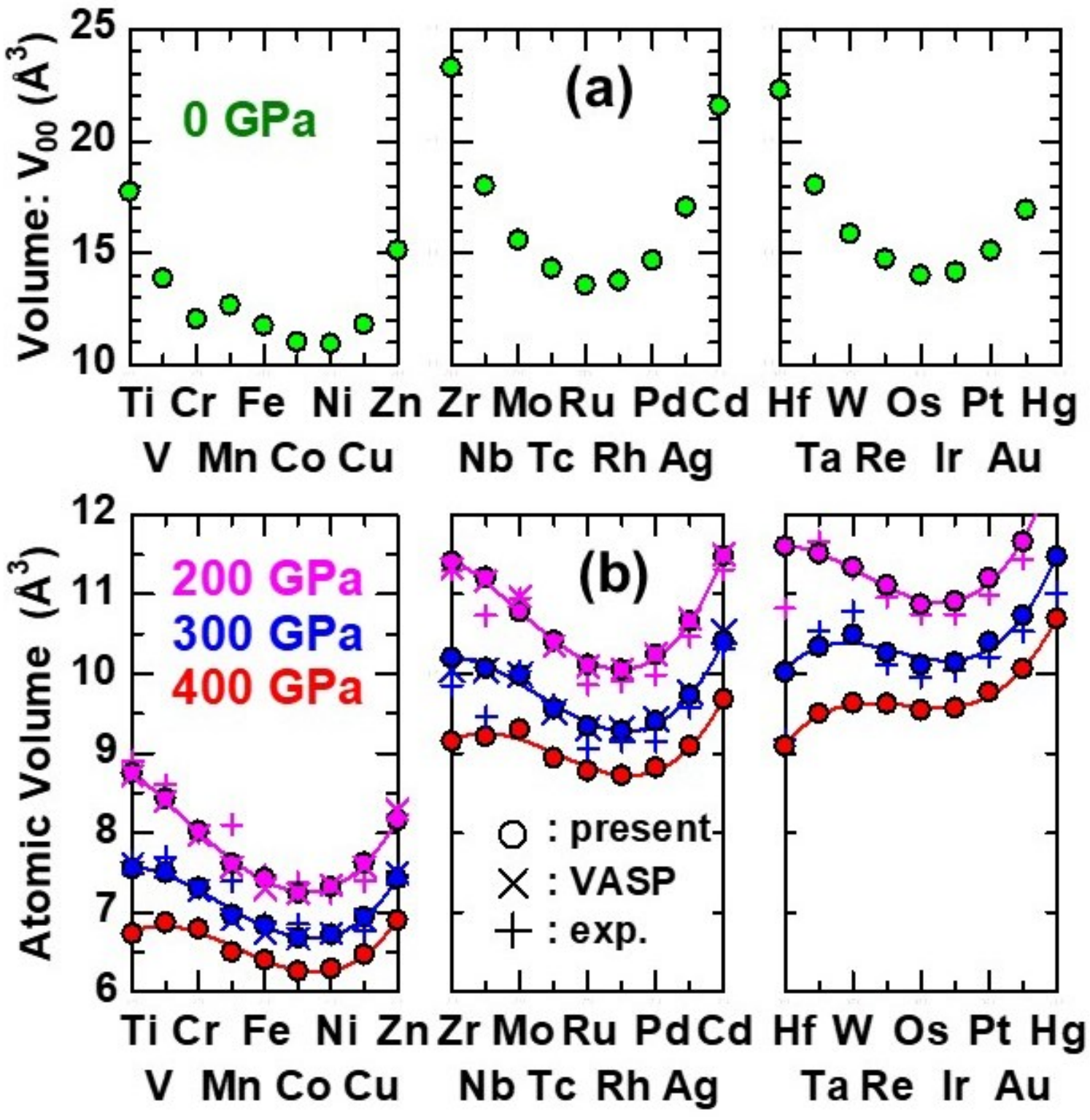

Fig. 1. (color online) Atomic volume of transition metals at 0, 200, 300, and 400 GPa (a, b). $V_{00}$ denotes the atomic volume under ambient conditions. Data points are taken from Table SII. The symbols + and × correspond to previous experimental results[2)] and calculations performed using VASP,[13)] respectively. The solid lines, which represent cubic fits of the present calculated data, show good agreement with the calculation results. The parabolic trend observed at 0 GPa evolves into a cubic-like behavior above 200 GPa, owing to the higher compressibility of bcc transition metals.

### 3. Results and Discussion

In this study, multiple pseudopotentials were tested, including those with semi-core valence states that explicitly account for with semi-core electrons (spn, spfn, and dn).[8)] The pseudopotential that yielded atomic volumes closest to experimental data was adopted (see Fig. S1).[3)] Furthermore, for the transition metals containing *f*-electrons (Hf, Ta, W, Re, Os, Ir, Pt, and Au), calculations were performed using the rel-PBE pseudopotential, incorporating spin–orbit coupling (SOC) to account for relativistic effects. In the *P*–*V* relationships, the calculated atomic volumes were 1%–3% larger than those obtained using the standard PBE pseudopotential. This result can be considered a general consequence of relativistic effects. The final *P*-*V* relationships are shown in Fig. S2.[3)]

Magnetization was considered for the Co, Ni, and Fe. For fcc-Co, the total magnetization decreased sharply from 1.4 to 0.0 $\mu_B$/cell as pressure increased from 92 to 93 GPa, accompanied by discontinuous volume reduction of ~2.7% (from 8.45 to 8.22 Å$^3$). This pressure corresponds to the hcp-fcc transition pressure.[9)] The EOS of fcc-Co was calculated based on data above 93 GPa. In Ni, Stoner magnetism was gradually suppressed, decreasing from 0.67 $\mu_B$/cell at 0 GPa to 0.53 $\mu_B$/cell at 400 GPa. Magnetization did not disappear, which was consistent with previous results.[10,11] For hcp-Fe, the magnetization remained at 0.0 $\mu_B$/cell up to 400 GPa.

The obtained *P*–*V* relationship and EOS were compared with the experimental data (Fig. S2).[3] The Vinet equation[12] was used to estimate the EOS, and the calculation results are summarized in Table SII.[3] For bcc-Hf, the experimental EOS was recalculated using the Vinet equation based on the data above 64 GPa, where the bcc phase became stable. The pressure scale was also corrected.[2]

Figs. 1(a) and (b) show the atomic volumes at 0, 200, 300, and 400 GPa, alongside previous experimental data[2] and first-principles calculation results.[13] At ambient conditions, the parabolic behavior of $V_{00}$ arises from the balance between nearest-neighbor attraction owing to *d*-electron filling in the *t2g* bonding band (contraction) and valence electron repulsion (expansion). Above 300 GPa, the atomic volume decreased significantly, and this picture no longer applied. The atomic volumes at 300 and 400 GPa for the 3*d*, 4*d*, and 5*d* transition metals exhibited a systematic cubic-like trend, closely reproducing the experimental data.[2] In contrast, the atomic volumes of the *p*-block main-group elements in the fourth, fifth, and sixth periods increased monotonically with increasing valence electron number, driven by the Pauli exclusion principle and repulsion between electrons.[2] Therefore, the systematic behavior of transition metals differs fundamentally from that of the main-group elements.

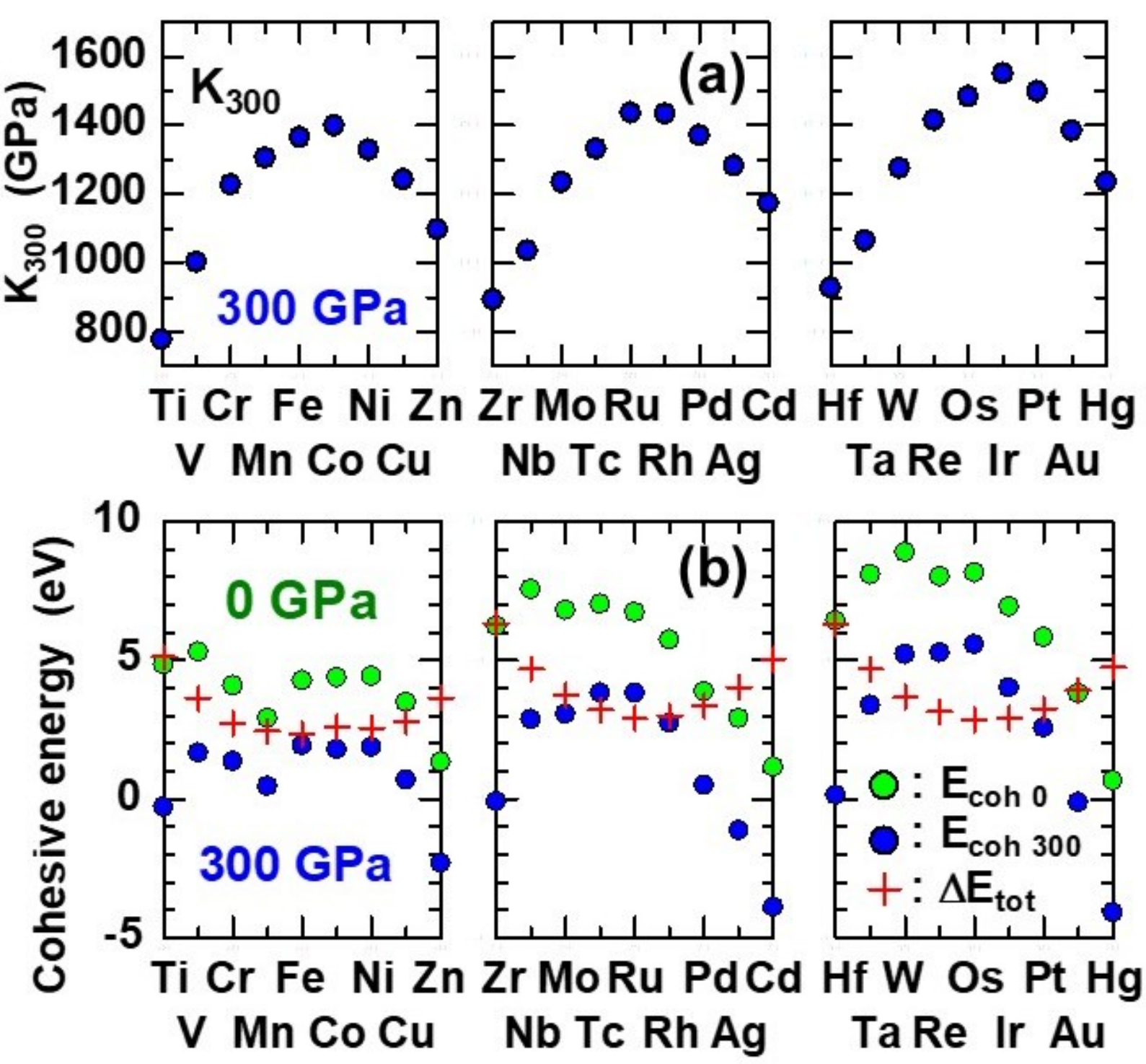

Fig. 2. (color online) (a) Bulk modulus ($K_{300}$) at 300 GPa and (b) cohesive energy ($E_{\mathrm{coh}}$) at 0 and 300 GPa together with the increase in total energy ($\Delta E_{\mathrm{tot}}$) due to compression up to 300 GPa for transition metals. Values of $E_{\mathrm{coh}300}$ at 300 GPa were estimated by subtracting $\Delta E_{\mathrm{tot}}$ from $E_{\mathrm{coh}0}$ at 0 GPa.[14] These values are summarized in Tables SII and SIII.[3]

For the bcc groups IV and V transition metals, the bulk modulus ($K_{300}$) at 300 GPa (Fig. 2(a)) was significantly smaller than that of the hcp groups VII and VIII metals. The atomic volumes of the former were more easily compressed than those of the latter, making the cubic-like trend of $V_{400}$ pronounced. Fig. 2(b) shows the cohesive energy ($E_{coh}$) at 0 and 300 GPa, along with the increases in total energy ($\Delta E_{tot}$) due to compression up to 300 GPa for transition metals. Values of $E_{coh300}$ at 300 GPa were estimated by subtracting $\Delta E_{tot}$ from $E_{coh0}$ at 0 GPa.[14] These values are summarized in Table SIII.[3] $\Delta E_{tot}$ for transition metals exhibited parabolic dependence on atomic number. For the bcc groups IV and V transition metals, $\Delta E_{tot}$ was larger than that of the hcp groups VII and VIII transition metals. This behavior was consistent with the trend observed in the bulk modulus. The large $\Delta E_{tot}$ values for bcc transition metals stem from their significant volume reduction, meaning that the external work ($P\Delta V$) under extreme compression up to 300 GPa is primarily stored as an increase in the total energy of the electronic system.

Under ambient conditions, transition metals exhibit a positive correlation between cohesive energy and bulk modulus, and a negative correlation between cohesive energy and atomic volume. Even under high compression at 300 GPa, the bulk modulus remained higher for elements with greater cohesive energy. However, in transition metals with a bcc structure, atomic volume decreased despite a significant reduction in cohesive energy. This unusual behavior is closely related to the *d*-electron band characteristics of the bcc structure.

The low bulk modulus, that is, the high compressibility of the bcc transition metals, is related to their valence electron state. Because bcc transition metals have relatively larger atomic volumes and fewer valence electrons, their valence electron density is lower than that of hcp or fcc transition metals. This low valence electron density contributes to their high compressibility, because the valence electron density is directly liked to degeneracy pressure under extreme compression. In addition, transition metals with the bcc structure consist solely of bonding orbitals that form the *d*-electron *t2g* band. The *t2g* band is a bonding orbital oriented along the [111] direction of the nearest- neighbor atoms in the bcc lattice. As the interatomic distance decreases, the overlap of the bonding orbitals increases, leading to widening of the *d*-band.[4] This enhanced overlap mitigates repulsive forces between these ion cores compared with hcp and fcc transition metals, where the antibonding *eg* band is occupied by *d*-electrons. Moreover, under ultrahigh pressure, the occupancy of the *t2g* band in the bcc structure increases further because of the *s*-*d* electron transition, reinforcing this effect.[4] The bonding electron charges also act as a screen for the nearest-neighbor ion cores, reducing repulsive interactions.

The present Quantum ESPRESSO calculations are consistent with previous DFT studies,[13] as both employed the same GGA–PBE pseudopotential. For fcc transition metals, the calculated values of atomic volume were larger than the experimental results, but were reproduced with an error margin of 2.5%. However, note that the experimental data for bcc (Nb, Hf) and hcp (Hg) metals significantly deviated from the calculated results, likely due to uncertainties in difference of the pressure scale and/or the extrapolation of the EOS curves estimated at pressures below 200 GPa. Therefore, additional experiments on these elements at pressures of up to 300 GPa are required.

Recently, Sakai *et al*. reported EOS curves for elements such as Cu, Re, and W based on X-ray diffraction experiments in this pressure range.[15] Their $V_{300}$ values are approximately 1% larger than those reported here, but the difference is remains within experimental uncertainty.

## 4. Conclusion

In this study, we determined the equations of state up to 400 GPa for 27 transition metals with bcc, hcp, and fcc structures using DFT calculations in Quantum ESPRESSO. The calculated atomic volumes were compared with previous experimental data[2] and first-principles results.[13] At multi-megabar pressures (300–400 GPa), the characteristic parabolic relationship between atomic volume and *d*-band electron number observed at ambient conditions transformed into a cubic-like behavior, consistent with experimental observations. This change arises from the higher compressibility of groups IV and V elements with the bcc structure compared to those with the hcp and fcc structures. The valence electron states of the bcc transition metals appear to play an important role in this behavior at ultra-high pressures above 300 GPa. Ourt results, based on atomic volume trends, provide new insights into the electronic origins of the behavior of transition metals at ultra-high densities. These findings offer fundamental data for studies of condensed matter physics under extreme conditions and contribute to the understanding of planetary interiors.

**Figure Caption**

Fig. 1. (color online) Atomic volume of transition metals at 0, 200, 300, and 400 GPa (a, b). $V_{00}$ denotes the atomic volume under ambient conditions. Data points are taken from Table SII. The symbols + and × correspond to previous experimental results[2)] and calculations performed using VASP,[13)] respectively. The solid lines, which represent cubic fits of the present calculated data, show good agreement with the calculation results. The parabolic trend observed at 0 GPa evolves into a cubic-like behavior above 200 GPa, owing to the higher compressibility of bcc transition metals.

Fig. 2. (color online) (a) Bulk modulus ($K_{300}$) at 300 GPa and (b) cohesive energy ($E_{\mathrm{coh}}$) at 0 and 300 GPa together with the increase in total energy ($\Delta E_{\mathrm{tot}}$) due to compression up to 300 GPa for transition metals. Values of $E_{\mathrm{coh300}}$ at 300 GPa were estimated by subtracting $\Delta E_{\mathrm{tot}}$ from $E_{\mathrm{coh0}}$ at 0 GPa.[14)] These values are summarized in Tables SII and SIII.[3)]

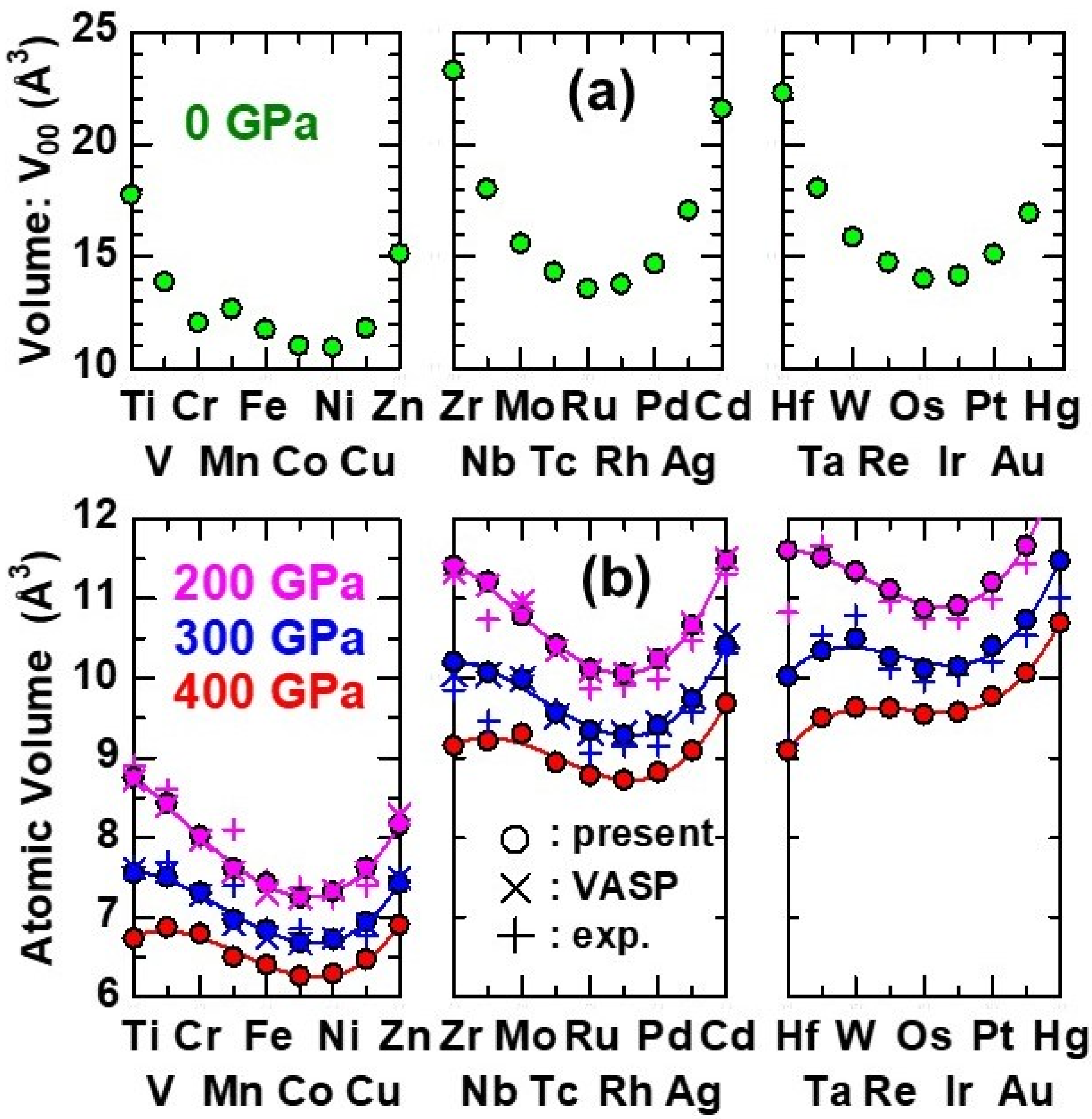

Fig. 1 by M. Geshi and Y. Akahama

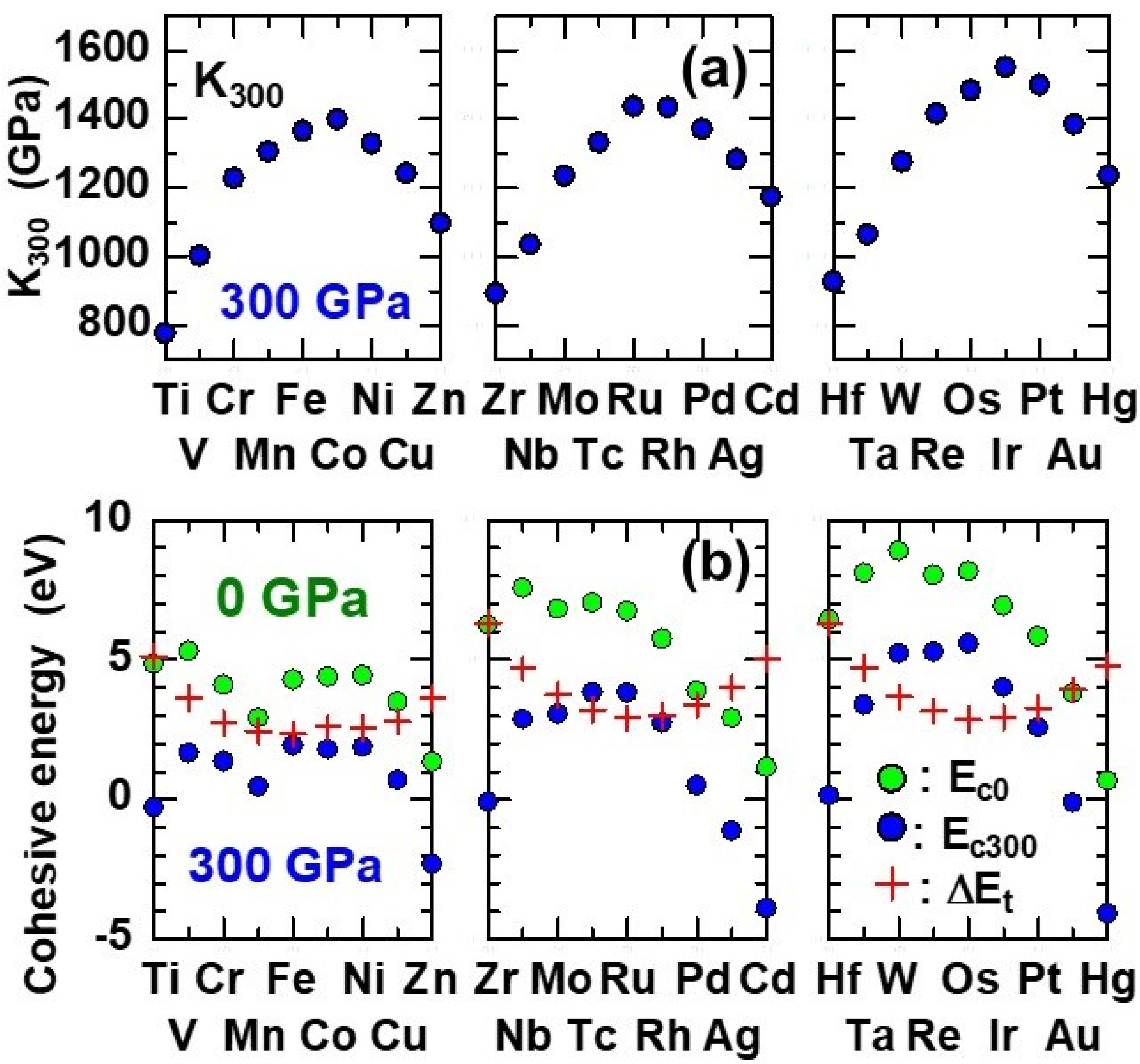

Fig. 2 by M. Geshi and Y. Akahama